\documentclass[onecolumn,showpacs,preprintnumbers,amsmath,amssymb]{revtex4}

\usepackage{graphicx}
\begin{document}

\title{Analytical study of non Gaussian fluctuations in a stochastic scheme of autocatalytic
reactions}

\author{Claudia Cianci}
\affiliation{Dipartimento di Sistemi e Informatica, University of Florence, Via
S. Marta 3, 50139 Florence, Italy}

\author{Francesca Di Patti}
\affiliation{Dipartimento di Fisica, Sapienza Universit\`{a} di
Roma, P.le A. Moro 2, 00185 Roma, Italy}

\author{Duccio Fanelli}
\affiliation{Dipartimento di Energetica, University of Florence and INFN, Via
S. Marta 3, 50139 Florence, Italy}

\author{Luigi Barletti}
\affiliation{ Dipartimento di Matematica, University of Florence
Viale Morgagni 67/A, 50134 Florence, Italy}

\begin{abstract}
A stochastic model of autocatalytic chemical  reactions is studied both numerically and analytically. The van Kampen perturbative scheme is
implemented, beyond the second order approximation, so to capture the non Gaussianity traits as displayed by the simulations. The method is targeted to the characterization of the third moments of the distribution of fluctuations, originating from a system of four populations in mutual interaction. The theory predictions agree well with the simulations, pointing to the validity of the van Kampen expansion beyond the conventional Gaussian solution.
\end{abstract}

\pacs{02.50.Ey, 05.40.-a, 82.20.Uv}

\maketitle

\vspace{0.8cm}

\section{Introduction}

The cell is a complex structural unit, that defines the building block of living systems \cite{alb02}.
It is made of by a tiny membrane, constituted by a lipid bilayer, which encloses a finite volume and protects
the genetic material stored inside. The membrane is semi-permeable: nutrients can leak in and serve
as energy storage to support the machinery functioning. Metabolism converts energy into
molecules, i.e. building cell components, and releases by-product.

Evolution certainly guided the ancient supposedly minimalistic cell entities, the so-called protocells \cite{mor88, dea86, lui06}, through subsequent steps towards the delicate and complex biological devices that we see nowadays. Focusing on primordial cell units, back at the origin of life, the most accredited scenario dictates that chemical reactions occurred inside vesicles, small cell-like structures in which the outer membrane takes the form of a lipid bilayer \cite{lui06}. Vesicles possibly defined the scaffold of prototypical cell models, while it is customarily believed that autocatalytic reactions might have been at play inside primordial protocell. The shared view is that protocell's volume might have been occupied by interacting families of replicators, organized in autocatalytic cycles. A chemical reaction is called autocatalytic if one of the reaction products is itself a catalyst for the chemical reaction. Even if only a small amount of the catalyst is present, the reaction may start off slowly, but will quickly develop once more catalyst is produced. If the reactant is not replaced, the process will again slow
down producing the typical sigmoid shape for the concentration of the product. All this is for a
single chemical reaction, but of greater interest is the case of many chemical reactions, where one
or more reactions produce a catalyst for some of the other reactions. Then the whole collection of
constituents is called an autocatalytic set. Autocatalytic reactions have been invoked in the context
of studies on the origin of life as a possible solution of the famous Eigen's paradox \cite{eigen}. This is a
puzzling logic concept which limits the size of self replicating molecules to perhaps a few hundred
base pairs. However, almost all life on Earth requires much longer molecules to
encode their genetic information. This problem is handled in living cells by the presence of
enzymes which repair mutations, allowing the encoding molecules to reach large enough sizes.
In primordial organisms, autocatalytic cycles might have contributed to the inherent robustness of the system,
translating in a degree of microscopic cooperation that successfully prevented the Eigen's evolutionary derive towards
self-destruction to occur. It is therefore of interest to analyze the coupled dynamics of chemicals
organized in extended cycles of autocatalytic reactions.

It is in this context that our work is positioned. We will in particular consider a model of
autocatalytic reactions confined within a bounded region of space. The model was pioneered by
Togashi and Kaneko \cite{togashi} and more recently revisited by \cite{dauxois, deanna}. It was in particular
shown that fluctuations stemming from the intimated discreteness of the scrutinized medium can seed a resonant effect
yielding to organized macroscopic patterns, both in time \cite{dauxois} and space \cite{deanna}.

As we shall clarify in the forthcoming discussion, the model here examined is intrinsically stochastic and
falls in the realm of the so called individual-based description. The microscopic dynamics follows
explicit rules governing the interactions among individuals and with the
surrounding environment. Starting from the stochastic scenario and performing a perturbative
development (van Kampen expansion \cite{vk}) with respect to a small parameter which encodes the amplitude of finite size fluctuations, one obtains,
at the leading order, the mean-field equations, i.e. the idealized continuum description for the concentration
amount. These latter govern in fact the coupled evolution of the average population
amount, as in the spirit of the deterministic representation. Including the next-to-leading order
corrections, one obtains a description of the fluctuations, as a set of linear
stochastic differential equations. Such a system can be hence analyzed exactly, so allowing us to
quantify the differences between the stochastic formulation and its deterministic analogue.
This analysis was performed in \cite{dauxois} with reference to the a-spatial version of model, and in
\cite{deanna} where the notion of space is instead explicitly included.

In this paper, we take one step forward by analytically characterizing the fluctuations beyond the second
order in the van Kampen perturbative scheme \cite{vk, gardiner}, i.e. the Gaussian approximation, and so quantifying higher
contributions in the hierarchy of moments of the associated distribution. As we shall demonstrate, and with reference to the
analyzed case study, we can successfully quantify non Gaussian fluctuations, within the van Kampen descriptive scenario,
in agreement with the recent investigations of Grima and collaborators \cite{Grima} and previous indications  of Risken and Vollmer \cite{ris}.

Again, let us emphasize that fluctuations do not arise from an external noise. Despite the evidence
that it is always present in actual population dynamics and that it is an essential ingredient of
life processes, noise is often omitted. When instead considered, it is frequently assumed to act as a
source of disorder and it is included in the dynamics as an external elements. At variance, the
individual-level approach allows us to investigate the unavoidable intrinsic noise, which originates
from the discreteness of the system and that has to be considered in any sensible model of natural
phenomena.

The paper is organized as follows. In the following section we will introduce the model under scrutiny.
Then we will turn to discussing the associated master equation, derive the mean field equation, and characterize the
fluctuations within the Gaussian approximation. Non Gaussian traits are revealed via numerical (stochastic) simulations
for small system sizes. These features are analytically inspected and explained in section VI by working in the framework of a generalized
Fokker-Planck formulation where the role of the finite population is explicitly accommodated for. Finally, in section VII
we sum up and conclude.

\section{The model}
\label{model}

The autocatalytic reaction scheme as introduced in \cite{dauxois}
describes the dynamics of $k$ species which interact according to
the following rules
\begin{eqnarray}
X_{i} + X_{i+1} &\stackrel {r_{i+1}}{\longrightarrow }&
2X_{i+1}\qquad
\text{with } X_{k+1} \equiv X_{1} \nonumber \\
E &\stackrel {\alpha_i}{\longrightarrow }& X_{i} \nonumber  \\
X_{i} &\stackrel{\beta_i}{\longrightarrow }& E
\label{reactions}
\end{eqnarray}
where $X_i$ denotes an element of the $i$--th species, while $E$ is
the null constituent or vacancies. The parameter  $r_{i}$ (with $r_{k+1} \equiv
r_{1}$) is the autocatalytic process rate, while $\alpha_{i}$ and
$\beta_{i}$ are the rates at which the molecules appear and
disappear from the system.  The size of the system is denoted by
$N$, then $\sum^{k}_{i=1} n_{i} + n_{E} = N$, where $n_E$ is the
number of $E$.

It is worth emphasizing that the concept of vacancies $E$ enables us to accommodate for
a finite carrying capacity of the hosting volume. The approach can be readily extended
to the case where the space is accounted for by formally dividing the volume in small
patches, each being characterized by a limited capacity. Species can then migrate between neighbors
cells, therefore visiting different regions of the spatial domain in which they are confined. This generalization is discussed in
\cite{deanna}. We will here solely consider the a-spatial version of the model, aiming at characterizing the fluctuations beyond
the canonical Gaussian approximation. In the following, we will introduce the master equation that rules the stochastic dynamics of the system
defined by the closed set of chemical equations (\ref{reactions}).

\section{The master equation and its expansion}
\label{intro} Let us start by introducing the master equation that
governs the evolution of the stochastic system described above.
First, it is necessary to write down the transition rates
$T(\boldsymbol{n}'|\boldsymbol{n})$ from the state $\boldsymbol{n}$
to the state $\boldsymbol{n}'$, where $\boldsymbol{n} \equiv
(n_{1},\ldots,n_{k})$ is the vector whose components define the
number of elements of each species at time $t$. These transition
rates are
\begin{eqnarray*}
& & T(n_{1},\ldots,n_{i}-1,n_{i+1}+1,\ldots,n_{k}|\boldsymbol{n}) =
r_{i+1} \frac{n_i}{N} \frac{n_{i+1}}{N}\,,  \\
& & T(n_{1},\ldots,n_{i}+1,\ldots,n_{k}|\boldsymbol{n}) =
\alpha_{i} \left( 1 - \frac{\sum^{k}_{j=1} n_j}{N} \right)\,,  \\
& & T(n_{1},\ldots,n_{i}-1,\ldots,n_{k}|\boldsymbol{n}) = \beta_{i}
\frac{n_i}{N}\,.
\end{eqnarray*}
In this way, the differential equation for the probability
$P(\textbf{n},t)$ reads
\begin{equation}\label{master}
\begin{split}
\frac{d}{dt}P(\textbf{n},t)=&\sum_{i=1}^{k}(\varepsilon_{i}^{+}
\varepsilon_{i+1}^{-})T(n_{1},...n_{i}-1,n_{i+1}+1,...n_{k})P(\textbf{n},t)
+\\&\sum_{i=1}^{k}(\varepsilon_{i}^{-}-1)T(n_{1},...,n_{i}+1,n_{i+1},...,n_{k})P(\textbf{n},t)
+\\&\sum_{i=1}^{k}(\varepsilon_{i}^{+}-1)T(n_{1},...,n_{i}-1,n_{i+1},...,n_{k})P(\textbf{n},t)
\end{split}
\end{equation}
where $\varepsilon_{i}^{\pm}$ are the step operators which act on an arbitrary  function $f(\mathbf{x})$ as $\varepsilon_{i}^{\pm}f(\mathbf{x}) = f(\ldots, x_i\pm 1, \ldots)$.

The above description is exact: no approximations have yet been
made. At this stage we could resort to numerical
simulations of the underlying chemical reactions by means of the Gillespie
algorithm~\cite{gil76,gil77}. This method produces realizations of
the stochastic dynamics which are formally equivalent to those found
from the master equation (\ref{master}). Averaging over many
realizations enables us to calculate quantities of interest. We will
comment on the results of such simulations, in the following. A different route is however possible
which consists in drastically simplifying the master equation, via a perturbative calculation,
the van Kampen system size expansion~\cite{vk, gardiner}. It is
effectively an expansion in powers of $N^{-1/2}$, which to the leading
order ($N \to \infty$) gives the deterministic equations describing
the system, while at next-to-leading order returns the finite $N$
corrections to these. The method consists in putting forward the ansatz:
\begin{equation}\label{ip}
\frac{n_{i}}{N}=\phi_i+\frac{\xi_{i}}{\sqrt{N}}.
\end{equation}
where $\xi_i$ is the $i$-th component of the $k$-dimensional stochastic variable  $\boldsymbol{\xi}=(\xi_1,\xi_2,...)$.
To proceed in the analysis we make  use of the working ansatz (\ref{ip}) into the
master equation (\ref{master}). Then, it is straightforward to show
that the operator $\varepsilon_i^{\pm}$ can be approximated as:
\begin{equation*}
\varepsilon_{i}^{\pm}=1\pm\frac{1}{N^{1/2}}
\frac{\partial}{\partial\xi_{i}} +\frac{1}{2N}
\frac{\partial^{2}}{\partial\xi_{i}^{2}}
\pm\frac{1}{3!N^{3/2}}\frac{\partial^{3}}{\partial\xi_{i}^{3}} + \ldots
\end{equation*}

The first step in the perturbative calculation consists in
expliciting in the master equation  the dependence on the
concentration vector $\textbf{y}=\textbf{n}/N$. It is legitimate to
assume that this latter quantity changes continuously with time, as
far as each instantaneous variation is small when compared to the
system size. We therefore proceed by defining the following
distribution:
\begin{equation*}
\Pi(\boldsymbol{\xi},t)=P(\textbf{y},t)= P\Big(\boldsymbol{\phi}(t)
+\frac{\boldsymbol{\xi}}{\sqrt{N}},t\Big).
\end{equation*}
A simple manipulation yields to:
\begin{equation*}
\frac{\partial P}{\partial t}=-\sqrt{N}\sum_{i=1}^{k}\frac{\partial
\Pi}{\partial \xi_i}\frac{d\phi_i}{dt}+\frac{\partial\Pi}{\partial
t}.\end{equation*}

Similarly one can act on the right  hand side of Eq. (\ref{master})
and hierarchically organize the resulting terms with respect to
their $N$--dependence. The outcome of such algebraic calculation are
reported in the following. We will in particular limit our
discussion to the Gaussian approximation, by neglecting, at this
stage, the $N^{-3/2}$ terms. We will then return on this important
issue and discuss the specific role that is played by $N^{-3/2}$
corrections.

\subsection{The $N^{-1/2}$ terms}
As concerns the terms of order $N^{-\frac{1}{2}}$ one obtains:
\begin{equation*}
-\frac{1}{\sqrt{N}}\sum_{i=1}^k
\frac{\partial\Pi}{\partial\xi_i}\frac{d\phi_i}{d\tau}=
\frac{1}{\sqrt{N}}\sum_{i=1}^k(r_{i+1} \phi_{i}\phi_{i+1} - r_i
\phi_{i-1}\phi_{i})\frac{\partial\Pi}{\partial
\xi_i}+\frac{1}{\sqrt{N}} \sum_{i=1}^k \Big[\beta_{i}\phi_i -
\alpha_{i}\Big(1-\sum_{m=1}^k \phi_m\Big)
\Big]\frac{\partial\Pi}{\partial \xi_i}
\end{equation*}
where the rescaled time $\tau$ is defined as $\tau= t/N$. Thus the
following system of differential equations holds for the
concentration amount $\phi_i$
\begin{eqnarray}\label{eq:mf_nnsp}
\frac{d \phi_i}{d\tau}& = &r_i\phi_{i-1} \phi_i - r_{i+1}\phi_i
\phi_{i+1} + \alpha_{i}\left(1-\sum_{m=1}^k  \phi_m\right) -
\beta_{i}\phi_i, \label{eq:meanfield}
\end{eqnarray}
which in turn corresponds to working within the so--called mean
field approximation and eventually disregard finite size
corrections.  We should emphasize that Eqs. (\ref{eq:mf_nnsp}) are
obtained by elaborating on the exact stochastic chemical model and
exploring the limit for infinite system size $N\rightarrow\infty$.

To make contact with previous  investigations \cite{dauxois} we shall assume the
simplifying setting with $\beta_i=\beta$, $\alpha_i=\alpha$  and
$r_i=r$ $\forall i$. Under this condition, all species
asymptotically converge to the fixed point $\phi^*$ which is readily
calculated as:
\begin{equation}\label{puntfix}
\alpha\Big(1- \sum_{m=1}^{k} \phi^*\Big)- \beta\phi^* =0 \qquad
\longrightarrow \qquad \phi^* = \frac{\alpha}{k\alpha + \beta}.
\end{equation}

\begin{center}
\begin{figure}[tb]
\includegraphics[scale=0.3]{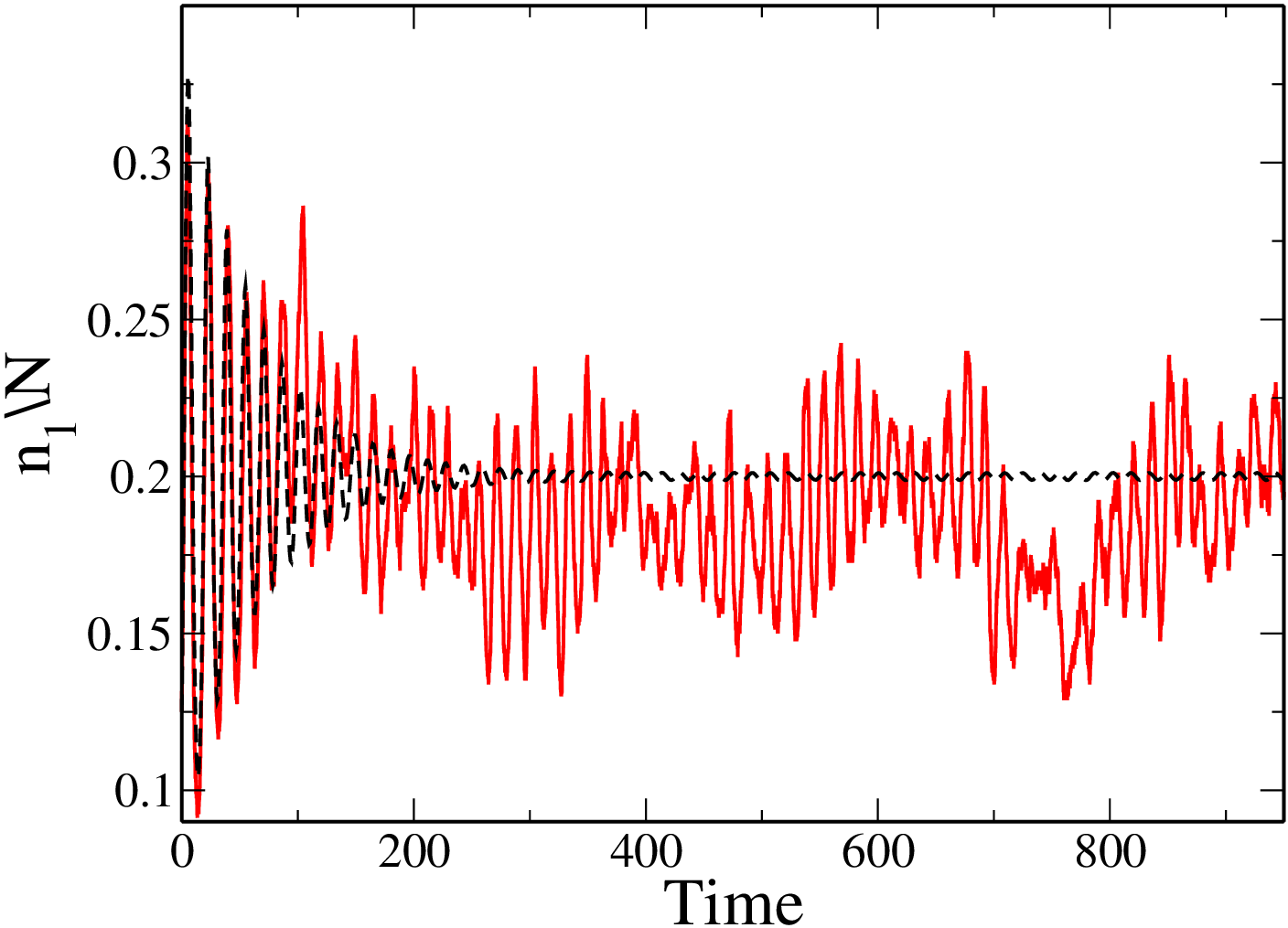}
\caption{Temporal evolution of one of the species concentrations
for a system composed by $4$ species and parameters set as $N=8190$,
$r_i=10$ and  $\alpha_i=\beta_i=1/64$ $\forall i$. The noisy (red
online) line represents one stochastic realization thought the
Gillespie algorithm \cite{gil76,gil77}, while the dashed black line
shows the numerical solution of the deterministic system given by
Eq. (\ref{eq:mf_nnsp}). \label{fig:mean_field}}
\end{figure}
\end{center}

We now turn to numerical simulation based on the Gillespie algorithm and discuss the case with $k=4$
species. As reported in Fig. \ref{fig:mean_field}, once the
initial transient has died out, the numerically recorded time series
keep on oscillating around the reference value as specified by
relation (\ref{puntfix}). The mean field dynamics has conversely
relaxed to the deputed equilibrium value. These oscillations stem
from the finite size corrections to the idealized mean field
dynamics and will be inspected in the following. We will be in
particular concerned with characterizing the statistical properties
of the observed signal, and quantify via rigorous analytical means
the moments of the distribution of the fluctuations.

\subsection{The $N^{-1}$ corrections}
Finite size effects  related to the  $N^{-1}$ corrections result in
the Fokker-Planck equation:
\begin{equation}\label{fp}
 \frac{\partial \Pi}{\partial \tau} = -\sum_{i=1}^k \frac{\partial}{\partial \xi_i}\Big[A_i(\xi) \Pi\Big] + \frac{1}{2} \sum_{j=1}^k \sum_{i=1}^k \frac{\partial^2}{\partial \xi_i \partial \xi_j} \Big[b_{ij} \Pi\Big]
 \end{equation}
which governs the  evolution of the distribution $\Pi(\cdot)$. Here
$A_i(\boldsymbol{\xi})$ reads:
\begin{equation*}
 A_i(\boldsymbol{\xi}) =  (r_i \phi_{i-1}-r_{i+1}\phi_{i+1})\xi_i  - r_{i+1}\phi_i \xi_{i+1} + r_i\phi_i\xi_{i-1} - \alpha_i\sum_{m=1}^k \xi_m - \beta_i\xi_i,
 \end{equation*}
 while $b_{ij}$ stands for the element  ${ij}$ of matrix $B$ defined as:
\begin{equation*}
 b_{ij} =\left\{ \begin{array}{ll}
 r_{i+1}\phi_i\phi_{i+1} + r_i\phi_i\phi_{i-1} +
 \alpha_i(1-\sum_{m=1}^k \phi_m) + \beta_i \phi_i & \quad \text{if} \quad  i=j\\
 -r_{i+1} \phi_{i} \phi_{i+1}                     & \quad \text{if} \quad  j=i+1\\
 -r_i \phi_{i-1}\phi_i                            & \quad \text{if} \quad  j=i-1\\
 0                                                & \quad \text{if} \quad |i-j|>1.
 \end{array}\right .
 \end{equation*}
For the sake of clarity we shall introduce the matrix $M$ of
elements $m_{ij}$ defined as:
\begin{equation*}
 m_{ij} =\left\{ \begin{array}{ll}
 r_i \phi_{i-1} - r_{i+1}\phi_{i+1}-
 \alpha_i- \beta_i            & \quad \text{if} \quad  i=j\\
 -r_{i+1} \phi_{i} - \alpha_i & \quad \text{if} \quad  j=i+1\\
 r_i \phi_i- \beta_i          & \quad \text{if} \quad  j=i-1\\
 - \alpha_i                   & \quad \text{if} \quad  |i-j| >1
 \end{array} \right.
 \end{equation*}
and so rewrite $A_i$ as:
\begin{equation*}
A_i=\sum_{j=1}^k m_{ij}\xi_j.
\end{equation*}

The Fokker-Planck equation (\ref{fp}) has been previously obtained in \cite{dauxois} and shown to
explain the regular oscillations displayed in direct stochastic simulations. The oscillations, in fact, materialize
in a peak in the power spectrum of fluctuations which can be analytically calculated working in the equivalent context of the
Langevin equation. Here, we take a different route and reconstruct the distribution of fluctuations through the calculation of the
associated moments.
 To allow for analytical  progress, we will assume again identical
chemical reactions rates for all species, namely $r_i = r$,
$\beta_i=\beta$ and $\alpha_i=\alpha $ $\forall i$. Moreover, we
will focus on the fluctuations around the equilibrium and so require
$\phi_i=\phi^*$ $\forall i$. Under these conditions the matrix $M$
is circulating and can be cast in the form:
\begin{equation*}
 M = \left [ \begin{array}{rrrrrr}
 m_0 & m_1 & m_2 &m_2 &  \ldots  & m_3\\
 m_3 & m_0 & m_1 &m_2 &  \ldots  & m_2\\
 m_2 & m_3 & m_0 & m_1 &\ldots  & m_2\\
 \ldots & \ldots & \ldots &\ldots & \ldots& \ldots\\
 m_1 & m_2 & m_2 &m_2 & \ldots & m_0\\
\end{array} \right].
 \end{equation*}
with $m_0= - \alpha- \beta$, $m_1 =-r\phi^* - \alpha $, $m_2 =-
\alpha $, and $m_3 =r\phi^* - \alpha $. The $k \times k$ matrix
reads instead:
 \begin{equation*}
 B= \left[ \begin{array}{rrrrrrr}
 b_0 & b_1 & 0 &  \ldots & 0 &b_1\\
 b_1 & b_0 & b_1 &  \ldots  &0 & 0\\
 0 & b_1 & b_0 &  \ldots  &0 & 0\\
 \ldots &  \ldots& \ldots& \ldots& \ldots & \ldots\\
 b_1 & 0 & 0 &  \ldots  &b_1 & b_0\\
 \end{array} \right]
 \end{equation*}
where $b_0= 2r\phi^* \phi^* + \alpha(1-k\phi^*) + \beta\phi^* $, and
$b_1 =- r\phi^* \phi^*$.

We recall that the solution of the Fokker--Planck equation
(\ref{fp}) is a multivariate Gaussian which is univocally
characterized by the associated families of first and second
moments. Working within this setting, it is hence sufficient to
derive the analytical equations that control the time evolution of the
first two moments of the distribution. We
will in particular provide closed analytical expressions for the asymptotic
moments and draw a direct comparison with the numerical
experiments.

\section{Analytical estimates of the fluctuations distribution moments}\label{parmom}
Define the moment of order $p$ for $\xi_i$ the quantity
\begin{equation*}
\langle\xi_{i}^{p}\rangle=\int \Pi(\boldsymbol{\xi})\xi_{i}^{p}d\boldsymbol{\xi}
\end{equation*}

Let us illustrate the analytical procedure  that is here adopted,
with reference to $\langle\xi_i^2\rangle$. To this end we start from Eq.
(\ref{fp}) and multiply it on both sides by the factor $\xi_i^2$.
Integrating over $\mathbb{R}^k$ in
$d\boldsymbol{\xi}=(d\xi_1,d\xi_2,...,d\xi_k)$, yields:
\begin{equation}\label{utile}
\int\xi_{i}^{2}\frac{\partial}{\partial \tau}\Pi(\boldsymbol{\xi},
\tau) d\boldsymbol{\xi}=\int
\xi_{i}^{2}\sum_{i}\frac{\partial}{\partial
\xi_{i}}A_{i}(\boldsymbol{\xi})\Pi(\boldsymbol{\xi},\tau)d\boldsymbol{\xi}
+\int\xi_{i}^{2}\frac{1}{2!}\sum_{i,j}\frac{\partial^{2}}{\partial
\xi_{i}\partial
\xi_{j}}b_{i,j}(\boldsymbol{\xi})\Pi(\boldsymbol{\xi},\tau)d\boldsymbol{\xi}.
\end{equation}

Consider the right hand  side of Eq. (\ref{utile}) and operate
two successive integrations by parts. Just two terms survive, as it
can be trivially argued for. Hence, bringing out the time derivative from the integral 
at the left hand side of Eq. (\ref{utile}), the sought equation for the second moments 
reads:
\begin{equation}\label{m21}
\dot{\langle\xi_{i}^{2}\rangle}=2m_{i,i}\langle\xi_{i}^{2}\rangle+2m_{i,
i-1}\langle\xi_{i}\xi_{i-1}\rangle+ 2m_{i, i+1}\langle\xi_{i}\xi_{i+1}\rangle+2m_{i, i+2}
\langle\xi_{i}\xi_{i+2}\rangle+b_{i,i}
\end{equation}
where $i=1,...,4$. Use has been  made of the definitions of the
coefficients $m_{ij}$. With analogous steps one immediately obtains
the differential equation that governs the time evolution of
quantity $\langle\xi_{i}\xi_{j}\rangle$:
\begin{equation}\label{m22}
\begin{split}
\dot{\langle\xi_{i}
\xi_{j}\rangle}=&m_{i,i}\langle\xi_{i}\xi_{i+1}\rangle+m_{i+1,i}\langle\xi_{i}^{2}\rangle+
m_{i,i+1}\langle\xi_{i+1}^{2}\rangle+m_{i,i+2}\langle\xi_{i+1}\xi_{i+2}\rangle\\&+m_{i+1,
i+1}\langle\xi_{i}\xi_{i+1}\rangle +m_{i+1,i+2}\langle\xi_{i}\xi_{i+2}\rangle
+m_{i+1,i}\langle\xi_{i}^{2}\rangle\\&+m_{i,i+3}\langle\xi_{i+3}\xi_{i+1}\rangle+b_{i,i+1}.
\end{split}
\end{equation}
which, in practice, encodes the degree of temporal correlation between
species $i$ and $j$. The picture is completed by providing the
equations for the first moments which read:
\begin{equation*}
\dot{\langle\xi_{i}\rangle}=m_{i,i}\langle\xi_{i}\rangle+m_{i, i-1}\langle\xi_{i-1}\rangle+
\langle\xi_{i}\rangle m_{i, i+1}-m_{i, i+2} \langle\xi_{i+2}\rangle.
\end{equation*}

Taking into account all possible permutations of the involved
indexes $i,j$, both ranging in the interval from $1$ to $4$, and
recalling the Eq.s (\ref{m21})--(\ref{m22}), one eventually
obtains a closed system of ten coupled ordinary differential
equations. For the simplified case $r_i=r$, $\alpha_i=\alpha$,
$\beta_i=\beta$ $\forall i$, this latter can be cast in a compact
form by introducing the matrix:
$$K=\left(\begin{array}{cccccccccc}
     2m_{0} & 2m_{1} & 2m_{2} & 2m_{3} & 0 & 0 & 0 & 0 & 0 & 0 \\
     m_{3} & 2m_{0} & m_{1} & m_{2} & m_{1} & m_{1} & m_{2} & m_{3} & 0 & 0 \\
     m_{2} & m_{3} & 2m_{0} & m_{1} & 0  & m_{1} & 0 & m_{2} & m_{3} & 0 \\
     m_{1} & m_{2} & m_{3} & 2m_{0} & 0  & 0 & m_{1} & 0 & m_{2} & m_{3} \\
     0 & 2m_{3} & 0 & 0 & 2m_{0}  & 2m_{1} & 2m_{2} & 0 & 0& 0 \\
     0 & m_{2} & m_{3} & 0 & m_{3}  & 2m_{0} &m_{1} & m_{1} & m_{2} & 0 \\
     0 & m_{1} & 0 & m_{3} & m_{2}  & m_{3} & 2m_{0} & 0 & m_{1} & m_{2} \\
     0 & 0 & 2m_{2} & 0 & 0  & 2m_{3} & 0 & 2m_{0} & 2m_{1}& 0 \\
     0 & 0 & m_{1} & m_{2} & 0 & m_{2} & m_{3} & m_{3} & 2m_{0} & m_{1} \\
     0& 0 & 0 & 2m_{1} & 0 & 0 & 2m_{2} & 0 & 2m_{3} & 2m_{0} \\
  \end{array}\right)$$
By further defining:
\begin{equation*}
\boldsymbol{X}=\left[ \langle\xi_{1}^{2}\rangle \quad \langle\xi_{1}\xi_{2}\rangle \quad
\langle \xi_{1}\xi_{3}\rangle \quad \langle\xi_{1}\xi_{4}\rangle \quad \langle\xi_{2}^{2}\rangle \quad
\langle \xi_{2}\xi_{3}\rangle \quad \langle\xi_{2}\xi_{4}\rangle \quad \langle\xi_{3}^{2}\rangle \quad
\langle \xi_{3}\xi_{4}\rangle \quad \langle\xi_{4}^{2}\rangle\right]
\end{equation*}
and the vector
\begin{equation*}
D=\left[ b_0 \quad b_1 \quad 0 \quad b_1 \quad b_0 \quad b_1 \quad 0
\quad b_0 \quad b_1  \quad b_0 \right]
\end{equation*}
one gets
\begin{equation*}
\dot{\boldsymbol{X}}=K\boldsymbol{X}+D \quad.
\end{equation*}

\begin{center}
\begin{figure}[tb]
\includegraphics[scale=0.4]{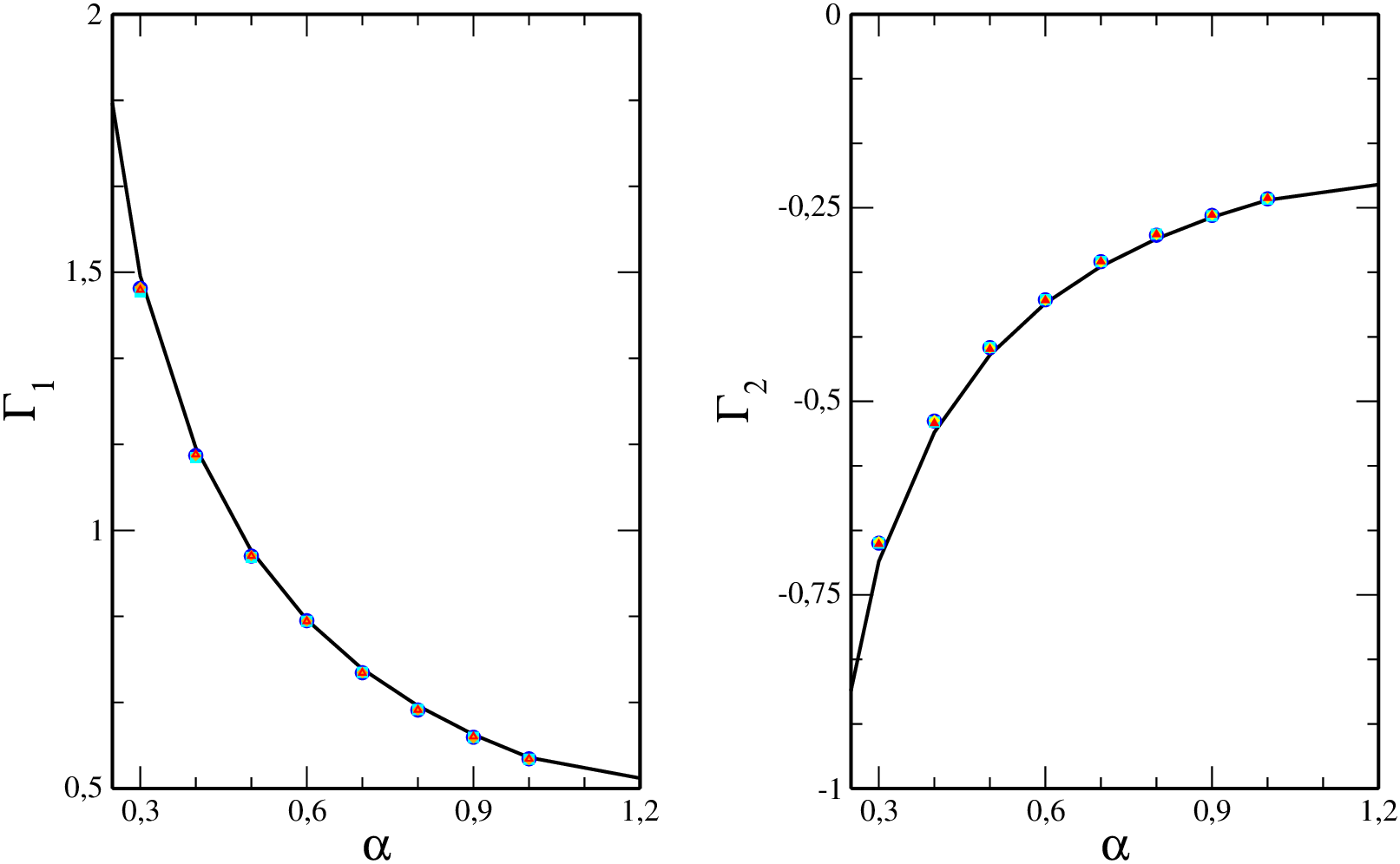}
\caption{Plots of the moments $\Gamma_1$ and $\Gamma_2$ as functions
of $\alpha$. The black lines show the theoretical predictions given
by Eq. (\ref{sol}), while the (colored online) symbols represent the
numerical simulations of the stochastic problem. Each symbol corresponds to a different
component of the family according to (\ref{simm}). Parameters are
set as $N=2000$, $\alpha=\beta$.\label{fig:gamma1gamma2}}
\end{figure}
\end{center}

\begin{center}
\begin{figure}[tb]
\includegraphics[scale=0.3]{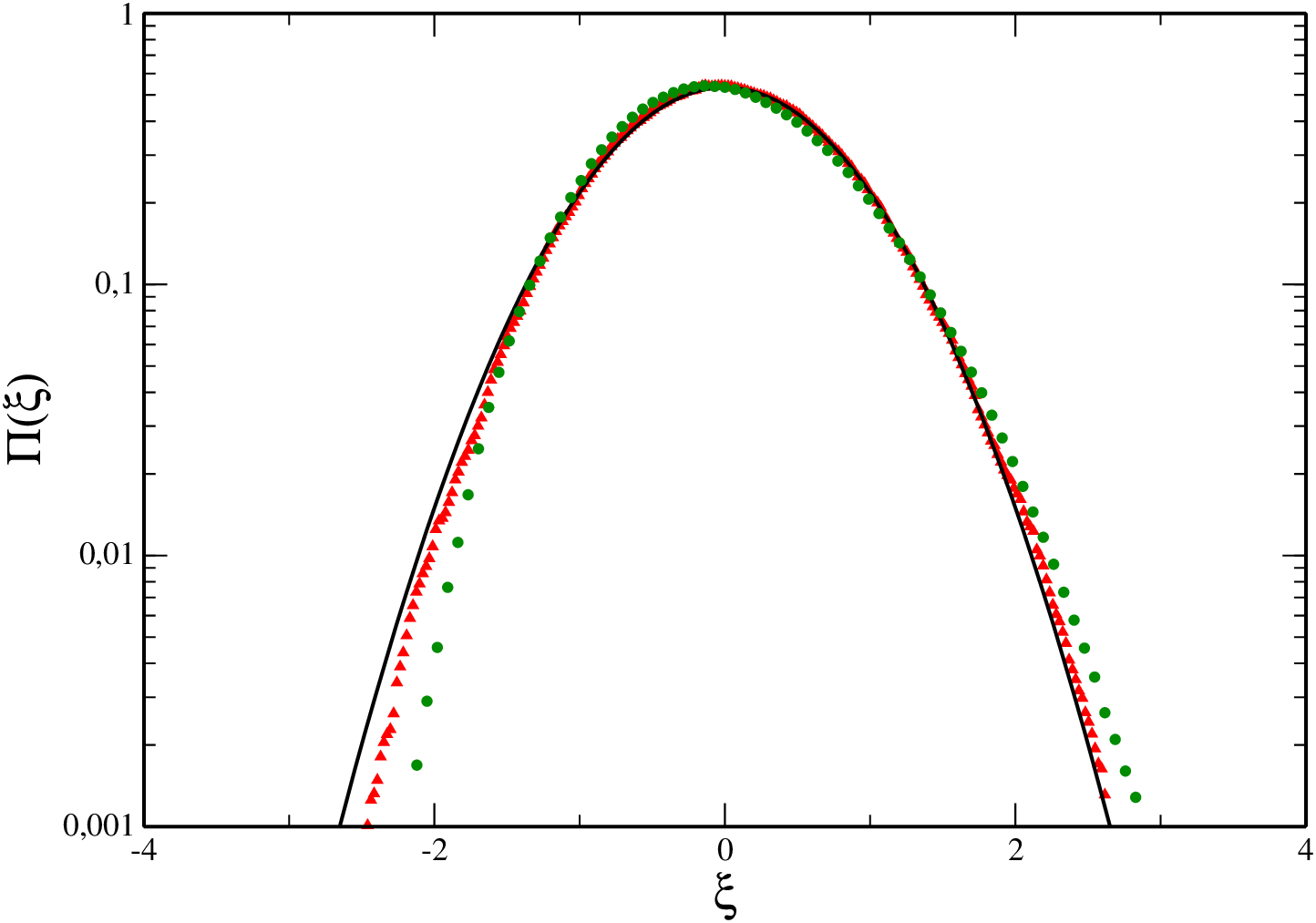}
\caption{Comparison between the stationary marginal Gaussian distribution and
the stochastic simulations (the y--axis has a logarithmic scale).
The solid (red on line) line shows the theoretical prediction
according to the van Kampen theory. The (green online) circles
represent the numerical distribution for a system with $N=200$,
while the (red online) triangles refer to a system with $N=2000$.
For all the curves $r=10$, $\alpha=\beta=0.1$.\label{fig:cupola}}
\end{figure}
\end{center}

As anticipated, we focus in particular on the late time evolution of the system, i.e. when the fluctuations'
distribution has converged to its asymptotic form. This request
translates into the mathematical condition $\dot{\boldsymbol{X}}=0$,
which implies dealing with an algebraic system of equations. Given
the peculiar structure of the problem, and by invoking a straightforward
argument of symmetry \footnote{It can be shown (see Fig. 1 of \cite{dauxois}) that the four families of chemicals evolve in pairs. Odd species $k=1,3$ are mutually
synchronized. The same applies to the even pairs $k=2,4$.}, one can identify  three families of independent unknowns,
namely:
\begin{equation}\label{simm}
\begin{split}
&\langle\xi_{1}^{2}\rangle=\langle\xi_{2}^{2}\rangle=\langle\xi_{3}^{2}\rangle=\langle\xi_{4}^{2}\rangle=:\Gamma_{1}\\&
\langle\xi_{1}\xi_{2}\rangle=\langle\xi_{1}\xi_{4}\rangle=\langle\xi_{2}\xi_{3}\rangle=\langle\xi_{3}\xi_{4}\rangle=:\Gamma_{2}\\&
\langle\xi_{1}\xi_{3}\rangle=\langle\xi_{2}\xi_{4}\rangle=:\Gamma_{3}.
\end{split}
\end{equation}
Closed analytical expressions for the  unknowns $\Gamma_1, \Gamma_2$
and $\Gamma_3$ as a function of the chemical parameters can be
derived and take the form:
\begin{eqnarray}
\Gamma_{1}&=&\frac{2b_{0}}{5\alpha}-\frac{b_{1}}{5\alpha}\nonumber \\
\Gamma_{2}&=&-\frac{b_{0}}{10\alpha}+\frac{3b_{1}}{10\alpha} \label{sol}\\
\Gamma_{3}&=&-\frac{b_{0}}{10\alpha}-\frac{b_{1}}{5\alpha} \nonumber
\end{eqnarray}
In deriving the above, we have  assumed a further simplifying condition, namely $\alpha=\beta$. The adequacy of the predictions is tested in Fig. \ref{fig:gamma1gamma2}, where $\Gamma_1$ and $\Gamma_2$ are plotted versus the independent parameter  $\alpha$. Recalling the explicit forms of $b_0$ and $b_1$ one can immediately appreciate that $\Gamma_3$ is indeed independent of $\alpha$. For this reason we here avoid to include $\Gamma_3$ in Fig. \ref{fig:gamma1gamma2}. One can moreover make use of the knowledge of the moments to reconstruct the profile of the distribution $\Pi(\boldsymbol \xi)$. In particular, and due to the symmetry of the model, we solely focus on the marginal distribution  $\Pi(\xi) = \Pi(\xi_i)$ for $i=1,\ldots,4$. In practice, we project the distribution in a one-dimensional  subspace by integrating over three out of four scalar independent variables $\xi_i$. In Fig. \ref{fig:cupola}, a comparison between theory and stochastic simulations (relative to small $N$ values) is drawn. While the agreement is certainly satisfying, deviations from the predicted Gaussian profile manifest as the population size shrinks. As we shall demonstrate, these distortions, which materialize in a skewed distribution, can be successfully explained  within an extended interpretative framework that moves from the van Kampen system size expansion.  In the following section we will hence extend the calculation beyond the Gaussian approximation. In doing so we will operate in the general setting for $\alpha \ne \beta$, but then specialize on the choice  $\alpha=\beta$ to drastically reduce the complexity of the inspected problem.

\section{Beyond the Gaussian approximation}\label{non_gauss}
We shall here go back to discussing the higher orders, $N^{-3/2}$
corrections to the Fokker-Planck equation. We will in particular
consider the various terms that contribute to the generalized Fokker-Planck equation grouping them as a function of the order of the
derivative involved.

The order $N^{-3/2}$ terms that involve the first derivatives can
be expressed as:
\begin{equation*}\begin{split}
\sum_{i=1}^{k}\Big[\frac{\partial}{\partial\xi_i}-\frac{\partial}{\partial\xi_{i+1}}\Big]r_{i+1}\xi_i\xi_{i+1}\Pi(\boldsymbol{\xi},\tau)&=\sum_{i=1}^{k}\frac{\partial}{\partial\xi_i}[r_{i+1}\xi_i\xi_{i+1}-r_i\xi_{i-1}\xi_{i}]\Pi(\boldsymbol{\xi},\tau)
\\&=-\sum_{i=1}^{k}\sum_{j=1}^{k}\frac{\partial}{\partial\xi_i}
l_{ij}\xi_i\xi_j\Pi(\boldsymbol{\xi},\tau)
\end{split}\end{equation*}
where $l_{ij}$ are the elements of the $k\times k$ circulant matrix
$L$ which, for $r_i=r \forall i$ reads:
\begin{equation*}
L=\left(\begin{array}{cccccc}
  0 & -r & 0 &\ldots &0 & r \\
  r & 0 & -r &\ldots &\ldots& 0 \\
  0 & r & 0 &\ldots &\ldots&\ldots\\
  \hdotsfor[2.0]{6}\\
\ldots& \ldots &\ldots&\ldots &-r& 0\\
  0 & \ldots &\ldots& r & 0&-r\\
 -r & 0 &\ldots& 0&r & 0\\
\end{array}
\right).
\end{equation*}

The order $N^{-3/2}$ contribution which depends on the second
derivatives can be also expressed in a matricial form. In fact, we
have:
\begin{multline*}
\frac{1}{2}\Big[\frac{\partial^2}{\partial\xi_i^2}\beta_i\xi_i-
\frac{\partial^2}{\partial\xi_i^2}\alpha_i\sum_w\xi_w+
r_{i+1}\frac{\partial^2}{\partial\xi_i^2}(\xi_i\phi_{i+1}+\phi_i\xi_{i+1})
+r_{i}\frac{\partial^2}{\partial\xi_{i}^2}(\xi_i\phi_{i-1}+\phi_i\xi_{i-1})
\\+2r_{i+1}\frac{\partial}{\partial\xi_i}\frac{\partial}{\partial\xi_{i+1}}(\xi_i\phi_{i+1}+\phi_i\xi_{i+1})
\Big]\Pi(\boldsymbol{\xi},\tau) =\sum_{i=1}^k\sum_{j=1}^k
\frac{\partial}{\partial\xi_i\partial\xi_j}d_{ij}
\end{multline*}
where the  $k\times k$ matrix $D$ of elements $d_{ij}$, for $r_i=r $ $\forall i$, reads:
\begin{equation*}
d_{ij}=\left\{
\begin{array}{ll}
\beta\xi_i-\alpha\sum_w\xi_w+r(\xi_i\phi^*+\xi_{i+1}\phi^*)+
r(\xi_i\phi^*+\xi_{i-1}\phi^*) & \quad \text{if} \quad  i=j\\
-r(\xi_i\phi^*+\xi_{i+1}\phi^*)& \quad \text{if} \quad  j=i+1\\
-r(\xi_i\phi^*+\xi_{i-1}\phi^*)& \quad \text{if} \quad  j=i-1\\
0 & \quad \text{otherwise}.
\end{array}
\right.
\end{equation*}

Finally, the third order derivatives contribute as:
\begin{equation*}
\Big[-\frac{1}{3!}\frac{\partial^3}{\partial\xi_i^3}\alpha_i(1-\sum_m\phi_m)
+\frac{1}{3!}\Big[\frac{\partial}{\partial\xi_i}-\frac{\partial}{\partial\xi_{i+1}}\Big]^{3}r_{i+1}\phi_i\phi_{i+1}
+\frac{1}{3!}\frac{\partial^3}{\partial\xi_i^3}\beta_i\phi_i\Big]\Pi(\boldsymbol{\xi},\tau)
=-\sum_{i,j,w}\frac{\partial^3}{\partial\xi_i\partial\xi_j\partial\xi_w}
e_{ijw}\Pi(\boldsymbol{\xi},\tau)
\end{equation*}
where, we introduced the matrix $E$ defined as:
\begin{equation*}
e_{ijw}=\left\{
\begin{array}{lll}
r(\phi^{*})^2                         & \quad j=i   &\quad w=i+1\\
r(\phi^{*})^2                         & \quad j=i-1 &\quad w=i-1\\
r(\phi^{*})^2                         & \quad j=i+1 &\quad w=i\\
-r(\phi^{*})^2                        & \quad j=i+1 &\quad w=i+1\\
-r(\phi^{*})^2                        & \quad j=i   &\quad w=i-1\\
-r(\phi^{*})^2                        & \quad j=i-1 &\quad w=i\\
-[\beta\phi^{*}-\alpha(1-k\phi^{*})]  & \quad i=j=w. & \phantom{i}
\end{array}
\right .
\end{equation*}

In conclusion, one gets the following equation  for the distribution
$\Pi(\boldsymbol{\xi},\tau)$ \cite{zamuner}:
\begin{multline}\label{fpgen}
\frac{\partial\Pi(\boldsymbol{\xi},\tau)}{\partial\tau}=
-\sum_{i=1}^{k}\frac{\partial}{\partial\xi_i}
[A_i(\boldsymbol{\xi})\Pi(\boldsymbol{\xi},\tau)]+\frac{1}{2}\sum_{i,j=1}^{k}
\frac{\partial^2}{\partial\xi_i\partial\xi_j}[b_{ij}\Pi(\boldsymbol{\xi},\tau)]
-\frac{1}{N^{1/2}}\sum_{i=1}^{k}
\frac{\partial}{\partial\xi_i}[C_i(\boldsymbol{\xi})\Pi(\boldsymbol{\xi},\tau)]\\
+\frac{1}{2N^{1/2}}\sum_{i,j=1}^{k}
\frac{\partial^2}{\partial\xi_i\partial\xi_j}[d_{ij}(\boldsymbol{\xi})\Pi(\boldsymbol{\xi},\tau)]
-\frac{1}{3!N^{1/2}}\sum_{i,j,w=1}^{k}\frac{\partial^3}{\partial\xi_i\partial\xi_j\partial\xi_w
}[e_{ijw}\Pi(\boldsymbol{\xi},\tau)]
\end{multline}
where:
$$C_i(\boldsymbol{\xi})=\sum_{j=1}^{k}l_{ij}\xi_i\xi_j.$$
We will refer to the latter as  to the \emph{generalized
Fokker--Planck equation}. In the following section we will discuss
the corrections to the Gaussian approximation as deduced by the
above mathematical framework.

\section{Non Gaussian corrections to the moments of the distribution}\label{momcal}
Starting from Eq.~(\ref{fpgen}) we  shall now assume $k=4$ and calculate the first three moments of the asymptotic distribution of the fluctuations
around the mean field equilibrium.  Clearly, the derivation can be in principle extended to evaluate the contribution of higher moments. The algebraic complexity of such an extension is however considerable and for this reason the third is the largest moment here characterized. The conclusions are nevertheless rather interesting as evaluating the third moment allows us to quantify the observed degree of skewness in the distribution of fluctuations.

When it comes to the first moment one gets:
\begin{equation}\label{eqmoprim}
\frac{d}{dt}\langle\xi_i\rangle = m_{i,i}\langle\xi_i\rangle+m_{i, i-1}\langle\xi_{i-1}\rangle+m_{i,
i+1}\langle\xi_i\rangle+m_{i, i+2}\langle\xi_{i+2}\rangle+\frac{1}{N^{1/2}}[l_{i,
i-1}\langle\xi_i \xi_{i-1}\rangle+l_{i, i+1}\langle\xi_i \xi_{i+1}\rangle]
\end{equation}
This equation differs from the one obtained in section \ref{parmom}
for the additional contribution
\begin{equation*}
\frac{1}{N^{1/2}}[l_{i, i-1}\langle\xi_i\xi_{i-1}\rangle+l_{i, i+1}\langle\xi_i
\xi_{i+1}\rangle]\quad .
\end{equation*}
Thanks to the symmetry of the system,  which ultimately stems from
having assumed $r_i=r$ $\forall i$, we can operate in a highly
simplified framework. We notice in fact that the above term is
function of the second moments, which have been estimated above and
quantified as
$\langle\xi_i\xi_{i+1}\rangle=\langle\xi_i\xi_{i-1}\rangle=\Gamma_2+o(1/\sqrt{N})$. Further
we observe that $l_{i,i-1}=-l_{i,i+1}$. Hence the corrections to the
Gaussian solution as exemplified in Eq. (\ref{eqmoprim})  contribute
with an overall term of order $N^{-3/2}$, which can be legitimately
neglected at this level of approximation. In conclusion the equation
for the first moments is identical to that obtained in section
\ref{parmom}.

Working in complete analogy, for the second moments we find:
\begin{equation*}
\langle\dot{\xi_i^{2}}\rangle=2m_{i,i}\langle\xi_i^2\rangle+2m_{i,
i+2}\langle\xi_i\xi_{i+2}\rangle+2m_{i, i+1}\langle\xi_i\xi_{i+1}\rangle +2m_{i,
i-1}\langle\xi_i\xi_{i-1}\rangle+b_{i,i} +\frac{2}{N^{1/2}}[l_{i,
i+1}\langle\xi_i^2\xi_{i+1}\rangle+l_{i, i-1}\langle\xi_i^2\xi_{i-1}\rangle]
\end{equation*}
for the variance of each involved species (recalling that the first moments are indeed null)
and
 \begin{eqnarray*}
\langle\dot{\xi_i\xi_{i+1}}\rangle &=& m_{i,i}\langle\xi_i\xi_{i+1}\rangle
+m_{i,i+2}\langle\xi_{i+1}\xi_{i+2}\rangle+m_{i, i+1}\langle\xi_{i+1}^2\rangle+m_{i+1,
i}\langle\xi_i^2\rangle + m_{i+1, i+1}\langle\xi_i\xi_{i+1}\rangle  \\
&\phantom{=}&+ m_{i+1,
i+2}\langle\xi_i\xi_{i+2}\rangle+m_{i,i+1}\langle\xi_i^2\rangle+m_{i,i+3}\langle\xi_{i+1}\xi_{i+3}\rangle
+\frac{1}{2}b_{i,i+1}+\frac{1}{2}b_{i+1,i} \\
&\phantom{=}&+\frac{1}{N^{1/2}}[l_{i, i+1}\langle\xi_i\xi_{i+1}^2\rangle+l_{i,
i-1}\langle\xi_i\xi_{i-1}\xi_{i+1}\rangle+l_{i+1, i}\langle\xi_i^2\xi_{i+1}\rangle+l_{i+1,
i+2}\langle\xi_i\xi_{i+2}\xi_{i+1}\rangle]
\end{eqnarray*}
for the mutual correlation between  distinct populations. The index
$i$ ranges from  $1$ to $4$. Again the extra contributions are
limited to the terms stored in square brackets and prove to be
negligible at this level of approximation. In fact the third order
correlations therein involved should scale as $N^{-1/2}$ as
requested by a simple consistency argument and as we shall prove a
posteriori. Then, also in this case, thanks to the specific form of
the matrix $L$, the additional contribution, stemming from  third
order moments, vanishes. We come hence to the conclusion that the
second moments are identical to those calculated in the preceding
section \ref{parmom} working within the Gaussian ansatz.

Let us now turn to calculating the third moments. After a lengthy
derivation we end up with:
\begin{multline*}
\langle\dot{\xi_i^3}\rangle=3m_0\langle\xi_i^3\rangle+3m_3\langle\xi_i^2\xi_{i-1}\rangle+3m_2\langle\xi_i^2\xi_{i+2}\rangle+3m_1\langle\xi_i^2
\xi_{i+1}\rangle+3b_0\langle\xi_i\rangle+\frac{3}{N^{1/2}}[m_4\langle\xi_i^2\rangle
\\+m_3\langle\xi_i\xi_{i+1}\rangle+m_3\langle\xi_i\xi_{i-1}\rangle +3m_2\langle\xi_i\xi_{i+2}\rangle]
+\frac{m_5}{N^{1/2}}+\frac{1}{N^{1/2}}[3r\langle\xi_i^3\xi_{i-1}\rangle-3r\langle\xi_i^3\xi_{i+1}\rangle]
\end{multline*}
where $m_0=-2\alpha$, $m_1=-\alpha-r/5$, $m_2=-\alpha$,
$m_3=-\alpha+r/5$, $m_4=2r/5$, $m_5=m_6=0$, $ m_7=-2r/5$, $m_8=r/25$
and $m_9=-r/5$. Here again, and as anticipated in the preceding
discussion, we have chosen the simplifying setting with
$\alpha=\beta$, which consequently implies $\phi^*=1/5$. Elaborating
on the symmetry one can identify five families of independent
moments, which obey to the above and the following differential equations:
\begin{eqnarray*}
\frac{d}{dt}\langle\xi_i^2\xi_{i-1}\rangle& =&
3m_0\langle\xi_i^2\xi_{i-1}\rangle+2m_3\langle\xi_i\xi_{i-1}^2\rangle
+2m_1\langle\xi_i\xi_{i+1}\xi_{i-1}\rangle +2m_2\langle\xi_i\xi_{i-1}\xi_{i+2}\rangle
+m_3\langle\xi_i^2\xi_{i+2}\rangle \\
&\phantom{=}&
+m_1\langle\xi_i^3\rangle+m_2\langle\xi_i^2\xi_{i+1}\rangle+b_0\langle\xi_{i-1}\rangle+2b_{1}\langle\xi_i\rangle\\
&\phantom{=}&
+\frac{1}{N^{1/2}}[m_3\langle\xi_{i-1}^2\rangle+m_6\langle\xi_i\xi_{i+1}\rangle+m_3\langle\xi_{i+1}\xi_{i-1}\rangle
+m_2\langle\xi_{i-1}\xi_{i+2}\rangle+m_7\langle\xi_i^2\rangle] \\
&\phantom{=}&-\frac{m_8}{N^{1/2}}
+\frac{1}{N^{1/2}}[-2r\langle\xi_{i-1}\xi_{i}^2\xi_{i+1}\rangle+2r\langle\xi_{i-1}^2\xi_{i}^2\rangle
+r\langle\xi_{i}^2\xi_{i-1}\xi_{i+2}\rangle-r\langle\xi_{i}^3\xi_{i-1}\rangle]
\end{eqnarray*}
and
\begin{eqnarray*}
\frac{d}{dt}\langle\xi_i^2\xi_{i+1}\rangle &=& 3m_0\langle\xi_i^2\xi_{i+1}\rangle+
2m_3\langle\xi_i\xi_{i-1}\xi_{i+1}\rangle+2m_1\langle\xi_i\xi_{i+1}^2\rangle
+2m_2\langle\xi_i\xi_{i+1}\xi_{i+2}\rangle\\
&\phantom{=}&+m_3\langle\xi_i^3\rangle+m_1\langle\xi_i^2\xi_{i+2}\rangle+m_2\langle\xi_i^2\xi_{i-1}\rangle
+b_0\langle\xi_{i+1}\rangle+2b_{1}\langle\xi_i\rangle\\
&\phantom{=}&+\frac{1}{N^{1/2}}[m_3\langle\xi_{i+1}^2\rangle+m_6\langle\xi_i\xi_{i+1}\rangle+m_3\langle\xi_{i+1}\xi_{i+3}\rangle
+m_2\langle\xi_{i+1}\xi_{i+2}\rangle+m_7\langle\xi_i^2\rangle]\\
&\phantom{=}&+\frac{m_8}{N^{1/2}}
+\frac{1}{N^{1/2}}[-2r\langle\xi_{i+1}^2\xi_{i}^2\rangle
+2r\langle\xi_{i-1}\xi_{i+1}\xi_{i}^2\rangle+r\langle\xi_{i}^3\xi_{i+1}\rangle-r\langle\xi_{i}^2\xi_{i+1}\xi_{i+2}\rangle]
\end{eqnarray*}
for adjacent populations with respect to the  assumed ordering. For
next--to--neighbors correlation one gets:
\begin{eqnarray*}
\frac{d}{dt}\langle\xi_i^2\xi_{i+2}\rangle &=&
3m_0\langle\xi_i^2\xi_{i+2}\rangle+2m_3\langle\xi_i\xi_{i-1}\xi_{i+2}\rangle+2m_1\langle\xi_i\xi_{i+1}\xi_{i+2}\rangle
+2m_2\langle\xi_i\xi_{i+2}^2\rangle+m_3\langle\xi_i^2\xi_{i+1}\rangle\\
&\phantom{=}&+m_1\langle\xi_i^2\xi_{i-1}\rangle+m_2\langle\xi_i^3\rangle +b_0\langle\xi_{i+2}\rangle
+\frac{1}{N^{1/2}}[-r\langle\xi_{i}^2\xi_{i+1}\xi_{i+2}\rangle+r\langle\xi_{i-1}\xi_{i+2}\xi_{i}^2\rangle]\\
&\phantom{=}&+\frac{1}{N^{1/2}}[m_2\langle\xi_{i+2}^2\rangle
+m_4\langle\xi_i\xi_{i+2}\rangle+m_3\langle\xi_{i-1}\xi_{i+2}\rangle+m_3\langle\xi_{i+1}\xi_{i+2}\rangle]
\quad .
\end{eqnarray*}
Finally, for correlations that involve three distinct species, we
find:
\begin{eqnarray*}
\frac{d}{dt}\langle\xi_i\xi_{i+1}\xi_{i-1}\rangle &=&
3m_0\langle\xi_i\xi_{i+1}\xi_{i-1}\rangle+m_3\langle\xi_{i+1}\xi_{i-1}^2\rangle+m_1\langle\xi_{i+1}^2\xi_{i-1}\rangle
+m_2\langle\xi_{i+2}\xi_{i+1}\xi_{i-1}\rangle\\
&\phantom{=}&+m_3\langle\xi_i^2\xi_{i-1}\rangle+m_1\langle\xi_i\xi_{i+2}\xi_{i-1}\rangle
+m_3\langle\xi_i\xi_{i+1}\xi_{i+2}\rangle +m_2\langle\xi_i\xi_{i-1}^2\rangle+m_1\langle\xi_{i}^2\xi_{i+1}\rangle\\
&\phantom{=}&+m_2\langle\xi_{i}\xi_{i+1}^2\rangle+b_1\langle\xi_{i-1}\rangle+b_{1}\langle\xi_{i+1}\rangle
+\frac{1}{N^{1/2}}[m_9\langle\xi_{i-1}\xi_i\rangle+m_9\langle\xi_{i+1}\xi_{i}\rangle
\\
&\phantom{=}&+m_7\langle\xi_{i-1}\xi_{i+1}\rangle]+\frac{1}{N^{1/2}}[-r\langle\xi_{i+1}^2\xi_{i}\xi_{i-1}\rangle+r\langle\xi_{i+1}\xi_{i}\xi_{i-1}^2\rangle].
\end{eqnarray*}

Clearly, the fourth moments enter  the equation for the third ones.
To close the system and so enable for quantitative predictions, we
can estimate the zero--th order contribution to the fourth moments
by recalling the Gaussian solution as obtained in Section
\ref{parmom} and neglecting the $1/\sqrt{N}$ terms. In formula:
\begin{eqnarray*}
\langle\xi_i^4\rangle&=&3(\langle\xi_i^2\rangle)^2\\
\langle\xi_i^3\xi_{j}\rangle&=&3\langle\xi_i^2\rangle\langle\xi_i\xi_j\rangle\\
\langle\xi_i^2\xi_{j}^2\rangle&=&\langle\xi_i^2\rangle\langle\xi_j^2\rangle+2(\langle\xi_i\xi_j\rangle)^2\\
\langle\xi_i^2\xi_{j}\xi_{k}\rangle&=&\langle\xi_i^2\rangle\langle\xi_j\xi_k\rangle+2\langle\xi_i\xi_j\rangle\langle\xi_i\xi_k\rangle.
\end{eqnarray*}
The above quantities can be  analytically estimated at equilibrium
and expressed as a function of respectively $\Gamma_1$, $\Gamma_2$,
$\Gamma_3$, as derived in section \ref{momcal}.

\subsection{The asymptotic evolution of the third moments}

Let us now write down the system of differential equations that
controls the dynamics of the five independent families of moments of
order three \footnote{The system reduces to five  families of independent moments as follows a the inherent symmetry of the problem to which we alluded in the preceding discussion.}. Such a system takes the form
\begin{equation}\label{eq:sistema}
\dot{\boldsymbol{X}}=V\boldsymbol{X}+S
\end{equation}
where $\boldsymbol{X}$ is
\begin{equation}\label{eq:vettoreX}
\boldsymbol{X}=\left[ \langle\xi_i^3\rangle \quad \langle\xi_i^2\xi_{i+1}\rangle \quad
\langle\xi_i^2\xi_{i-1}\rangle \quad \langle\xi_i^2\xi_{i+2}\rangle \quad
\langle\xi_i\xi_{i+1}\xi_{i-1}\rangle \right]
\end{equation}
and the matrix of coefficients $V$ reads:
\begin{equation*}
V=\left(
  \begin{array}{ccccc}
    3m_0 & 3m_1 & 3m_3 & 3m_2 & 0 \\
    m_3 & 3m_0 & 2m_1+m_2 & m_1 & 2m_3+2m_2 \\
m_1 & 2m_3+m_2 & 3m_0 & m_3 & 2m_1+2m_2 \\
 m_2 & m_3 & m_1 & 3m_0+2m_2 & 2m_3+2m_1 \\
    0 & m_2+m_1 & m_3+m_2 & m_3+m_1 & 3m_0+m_1+m_3+m_2 \\
  \end{array}
\right).
\end{equation*}
Finally the vector $S$ is:
\begin{equation*}
S=1/\sqrt{N}\left[s_1 \quad s_2 \quad s_3 \quad s_4 \quad s_5
\right]
\end{equation*}
where:
\begin{eqnarray*}
s_1 & = & 3m_4\Gamma_1+6m_3\Gamma_2+3m_2\Gamma_3\\
s_2 & = &
m_3\Gamma_1+m_7\Gamma_1+m_3\Gamma_3+m_2\Gamma_2+m_6\Gamma_2+
m_8+[-2r(\Gamma_1)^2+2r\Gamma_3\Gamma_1-2r\Gamma_3\Gamma_2+2r\Gamma_1\Gamma_2] \\
s_3 & = & m_3\Gamma_1+m_7\Gamma_1+m_3\Gamma_3+m_2\Gamma_2+
m_6\Gamma_2-m_8+[2r(\Gamma_1)^2-2r\Gamma_1\Gamma_3+2r\Gamma_3\Gamma_2-2r\Gamma_1\Gamma_2]
\\
s_4 & = &m_4\Gamma_3+2m_3\Gamma_2+m_2\Gamma_1 \\
s_5 & = &m_7\Gamma_3+2m_9\Gamma_2 \quad .
\end{eqnarray*}

\begin{center}
\begin{figure}[p]
\includegraphics[scale=0.4]{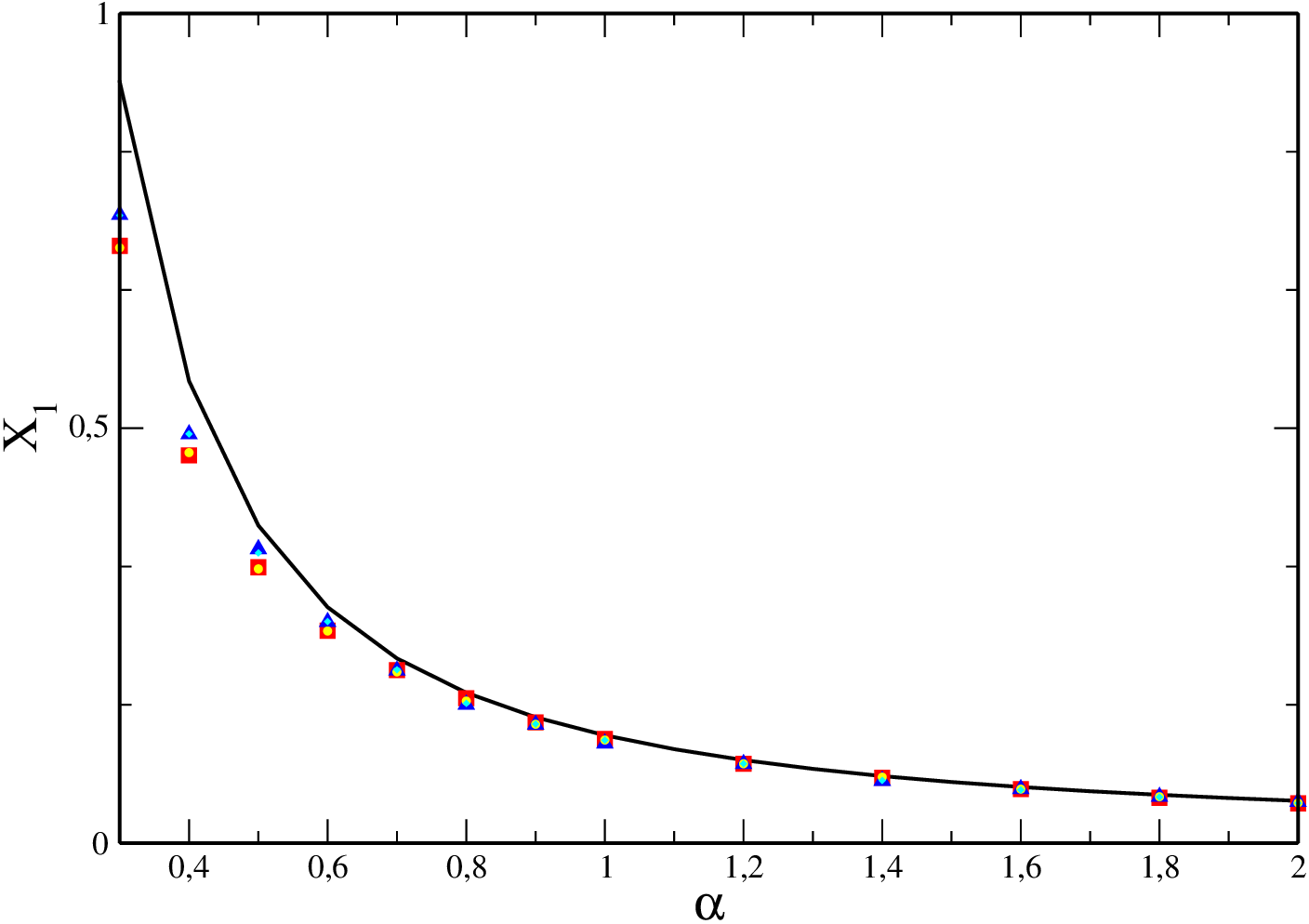}
\caption{Plots of  $X_1$ (see Eq.
(\ref{eq:vettoreX})) as functions of the parameter $\alpha$, for a
system with $\beta=\alpha$, $N=200$ and $r=10$. The solid black
lines represent the numerical solution of the system
(\ref{eq:sistema}), while the symbols refer to the stochastic
simulations (each of the four symbols is associated to a different
species).\label{m3_1}}
\end{figure}
\end{center}

\begin{center}
\begin{figure}[p]
\includegraphics[scale=0.4]{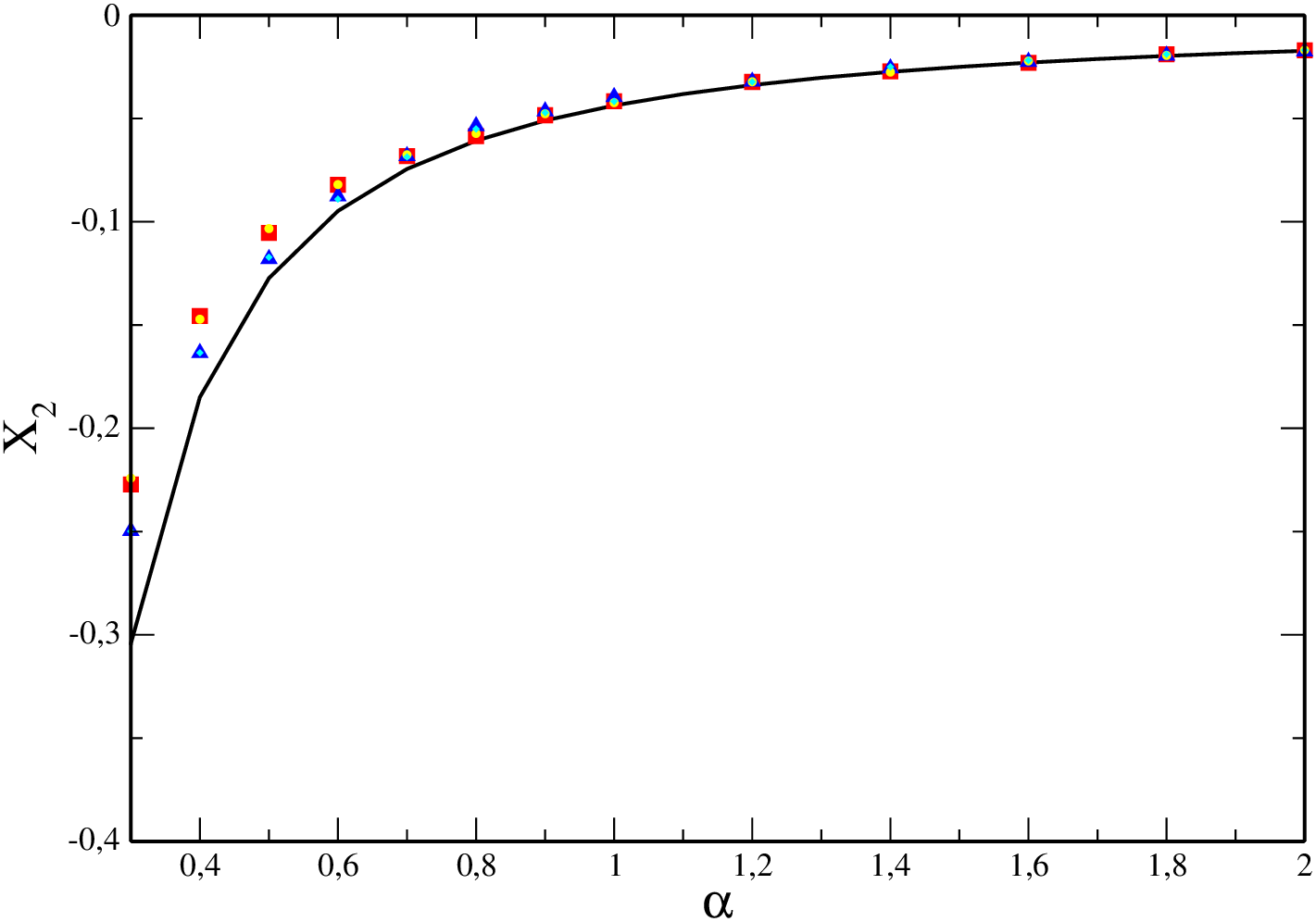}
\caption{Plots of  $X_2$  as functions of the parameter $\alpha$. For the parameters' setting and the explanation of the symbols see caption of Fig. \ref{m3_1}.\label{m3_2}}
\end{figure}
\end{center}

\begin{center}
\begin{figure}[tbp]
\includegraphics[scale=0.4]{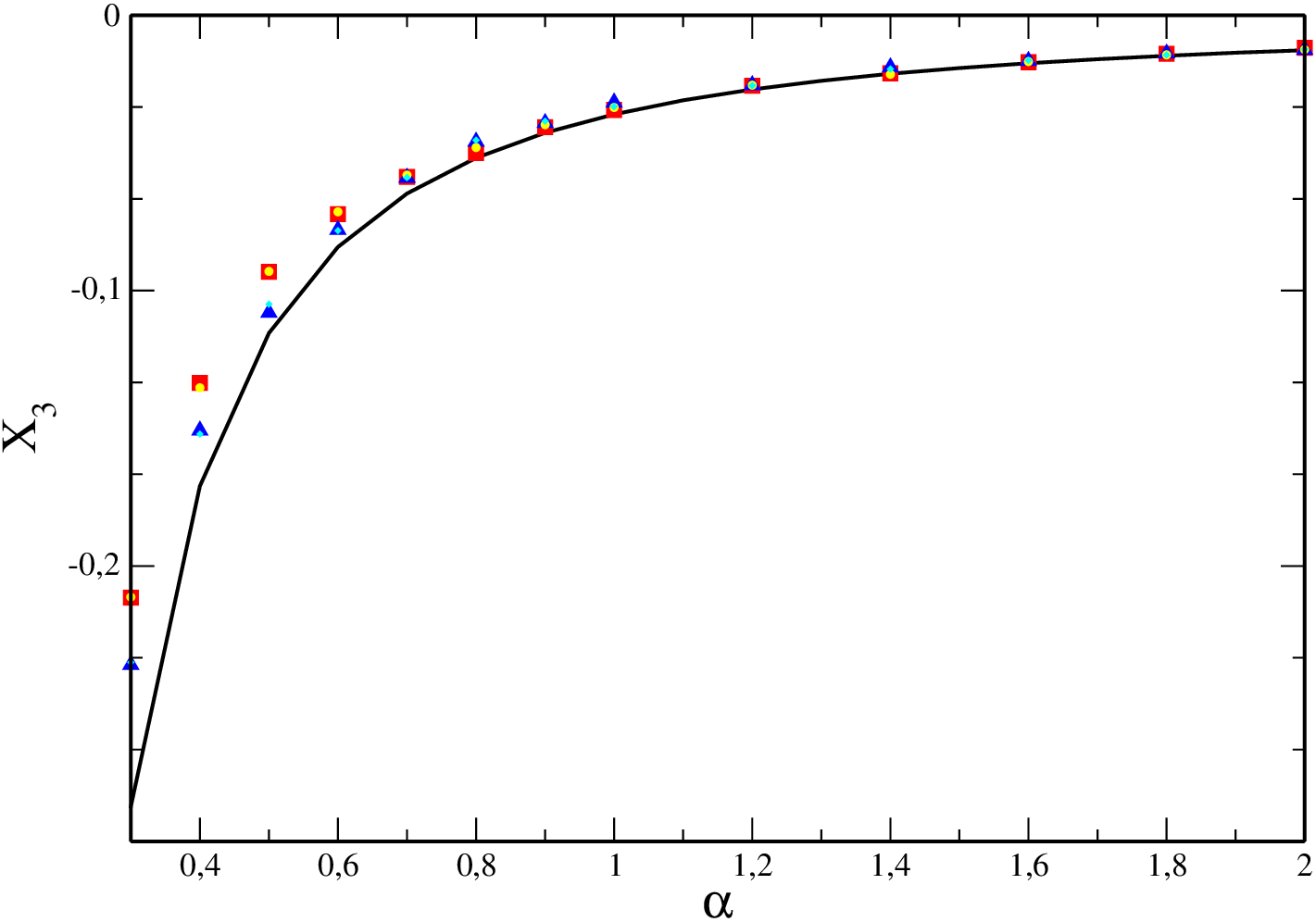}
\caption{Plots of  $X_3$ as functions of the parameter $\alpha$. For the parameters' setting and the explanation of the symbols see caption of Fig. \ref{m3_1}.\label{m3_3}}
\end{figure}
\end{center}

We now turn to numerical simulations to validate  the correctness of the theory. Stochastic simulations are performed for small
systems ($N=200$) and the time evolution of the third moments is monitored for each of the considered species and by varying the parameter
$\alpha$, while keeping $r$ unchanged. Results are displayed in Fig.s \ref{m3_1}--\ref{m3_5}, where the simulations outcome (symbols) are compared to the theory predictions. The agreement has to be considered satisfactory, a conclusion which a posteriori validates the theory assumptions and in particular confirms the predictive ability of the van Kampen expansion beyond the Gaussian approximation \cite{Grima, vk}.

\begin{center}
\begin{figure}[p]
\includegraphics[scale=0.4]{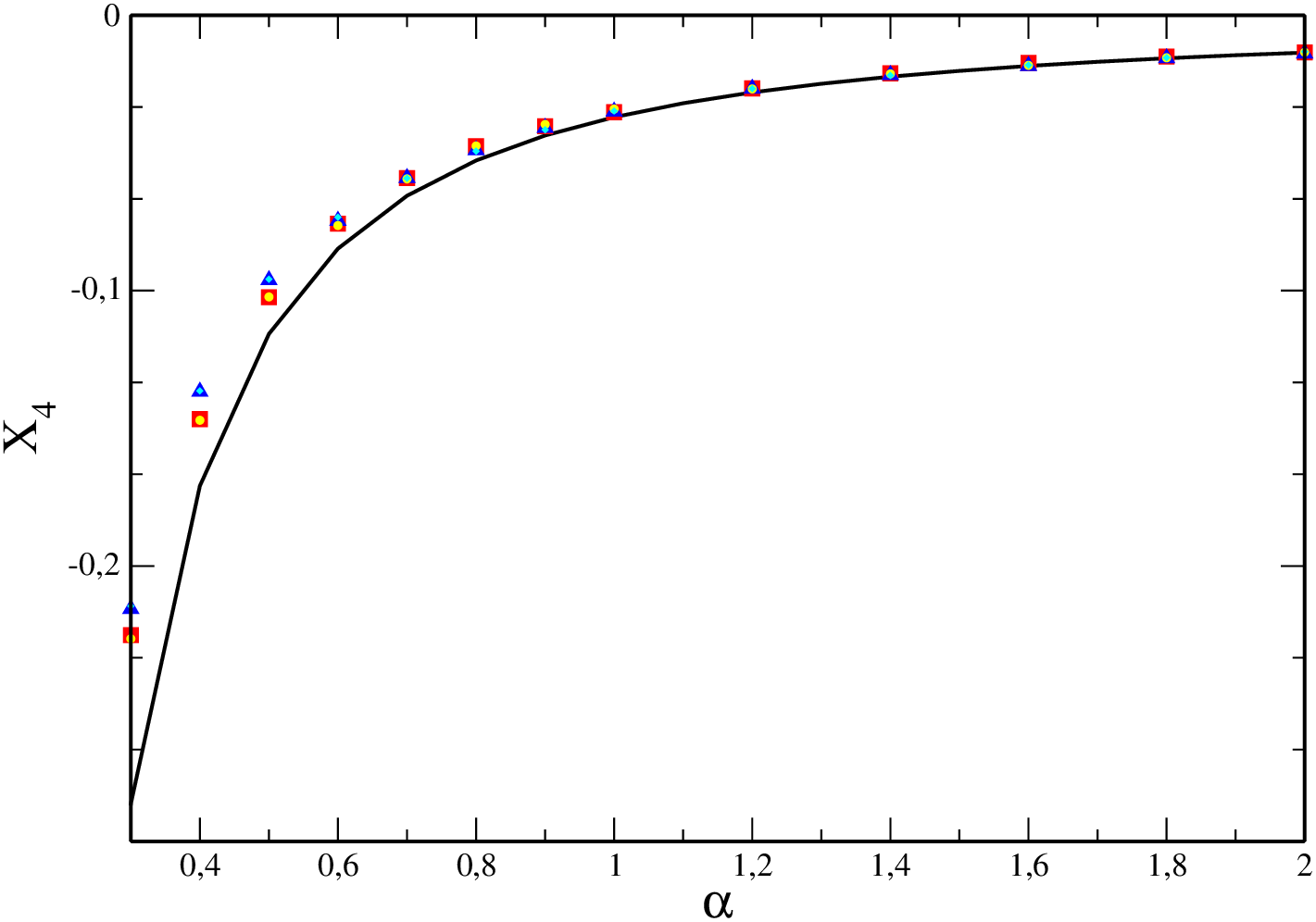}
\caption{Plots of $X_4$ as functions of the parameter $\alpha$. For the parameters' setting and the explanation of the symbols see caption of Fig. \ref{m3_1}.}
\end{figure}
\end{center}

\begin{center}
\begin{figure}[p]
\includegraphics[scale=0.4]{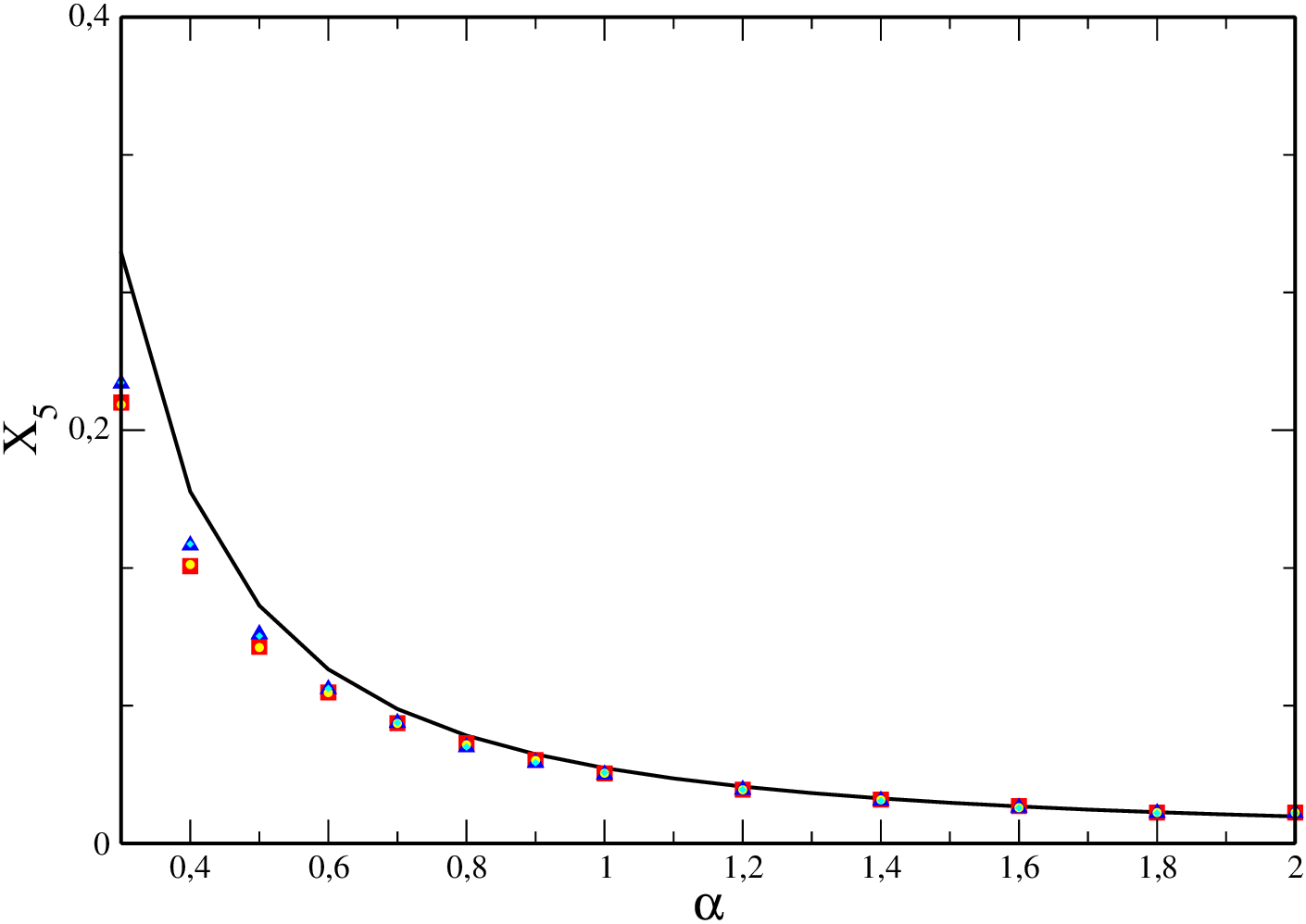}
\caption{Plots of $X_5$ as functions of the parameter $\alpha$. For the parameters' setting and the explanation of the symbols see caption of Fig. \ref{m3_1}.\label{m3_5}}
\end{figure}
\end{center}

\section{Conclusion}
The study of an extended set of autocatalytic reactions proves interesting in many respects. The system self-organizes at the macroscopic level, both in space and time, as follows a non linear resonance mechanism that enhances the stochastic fluctuations stemming from the  finite size. The spontaneous emergence of collective patterns, as well as regular time oscillations in such a system, was recently addressed \cite{dauxois, deanna} by analyzing in detail the underlying stochastic process via the celebrated van Kampen expansion, truncated at the Gaussian oder of approximation. In \cite{deanna}, it was also speculated that the intrinsic ability of the autocatalytic systems to drive self-organized structures might have played a role in the evolutionary selection of efficient cells, starting from minimalistic protocells entities. It was in fact argued that oscillatory, spatially extended patterns, might have resonate with the innate ability of a vesicle container to divide in two. One could imagine that the oscillations trigger the splitting event and thus favor a natural synchronization between the fission of the vesicle and the rate of production of the genetic material stored inside, which needs to be passed to the next generation offspring.

Besides these highly speculative considerations, which deserve to be carefully checked within a self-consistent picture, we are here interested in extending the perturbative calculation beyond the second order approximation and challenge its adequacy in capturing the deviation from the idealized Gaussian behavior. Recent support on the validity of the van Kampen higher orders calculation have been provided by Grima and collaborators \cite{Grima}.
We here bring one more evidence on the accuracy of the procedure within a rather complex model, where different species are simultaneously made to interact. Numerical simulations performed in a stochastic setting with modest sizes of the population involved, so to magnify the role played by
finite size corrections, confirm the correctness of the theory predictions. Due to the complexity of the proposed model, it is not possible to evaluate a large gallery of successive moments and so reconstruct the full distribution of fluctuations. The analysis is hence limited to the third moment, which however quantifies the degree of skewness of the recorded fluctuations. In a separate contribution \cite{cianci_voter}, we will
return on the issue of the validity of the van Kampen ansatz, working within a considerably simpler model that enables us to explicitly calculate all
the moments of the distribution at any order of the expansion. We are hence able to recover a general and exact analytical solution that, we anticipate, agrees very well with the simulations, inline with the conclusion of this work.

\end{document}